\documentstyle[12pt,epsfig]{article}
\newcommand{\lsim}{\,{\buildrel < \over {_\sim}}\,}
\newcommand{\gsim}{\,{\buildrel > \over {_\sim}}\,}

\begin{document}

\baselineskip 15pt
\parindent=15pt
\rightline{September 2000}
\rightline{JYFL-4/00}
\rightline{LPT Orsay 00-73}
\vspace{0.3cm}

\begin{center}
{\bf CONSTRAINTS FOR THE NUCLEAR SEA QUARK DISTRIBUTIONS FROM 
THE DRELL-YAN PROCESS AT THE SPS}\\ \vskip .5 truecm
{\bf K.J. Eskola$^{\rm a,d}$\footnote{kari.eskola@phys.jyu.fi},
V.J. Kolhinen$^{\rm a}$\footnote{vesa.kolhinen@phys.jyu.fi},
C.A. Salgado$^{\rm b}$\footnote{carlos.salgado@th.u-psud.fr} and 
R.L. Thews$^{\rm c}$\footnote{thews@physics.arizona.edu}
}\\

\vspace{1cm}
{\em $^{\rm a}$ Department of Physics, University of Jyv\"askyl\"a,\\
P.O.Box 35, FIN-40351 Jyv\"askyl\"a, Finland\\}

\vspace{0.3cm}

{\em $^{\rm b}$ Laboratoire de Physique 
Th\'eorique\footnote{Laboratoire associ\'e au Centre National de la Recherche
Scientifique - URA D0063.},Universit\'e de Paris XI,\\ B\^atiment 211,
F-91405 Orsay Cedex, France}

\vspace{0.3cm}

{\em $^{\rm c}$ Department of Physics, University of Arizona, \\
Tucson, Arizona 85721, USA}
\vspace{0.3cm}

{\em $^{\rm d}$ Helsinki Institute of Physics, \\
P.O. Box 9, FIN-00014 University of Helsinki, Finland}

\vspace{1cm}

\end{center}

\begin{abstract}
\noindent
Nuclear modifications to the Drell-Yan dilepton production cross sections in
p+$A$ and $A+A$ collisions in the leading twist
approximation are caused by nuclear effects in the parton distributions of
bound nucleons. For non-isoscalar nuclei, isospin corrections must also
be considered.  We calculate these effects 
for p+$A$ and Pb+Pb collisions at CERN SPS
energies. Our goal is to place constraints on nuclear effects
in sea quark distributions in the region $x\gsim 0.2$. We show
that the net nuclear corrections remain small for p+$A$ collisions at
$E_{\rm lab}=450$ GeV.  However, in Pb+Pb collisions at $E_{\rm
lab}=158\,A$GeV, effects of $\gsim 20\%$ are predicted at large
$M$.  The data collected by the NA50 collaboration could thus be used
to constrain the nuclear effects in the sea quark distributions in the
region of the EMC effect, $x\gsim 0.3$.

\end{abstract}
\vfill

\newpage

\section{Introduction}

In this work, we study perturbatively calculable QCD processes
in high energy p+$A$, $A+A$ and $A+B$ collisions.
In the search for the Quark-Gluon Plasma
(QGP) in ultrarelativistic heavy ion collisions, these inclusive hard
processes provide reference cross sections for detecting specific
signatures of the QGP, such as suppression of heavy quarkonia,
production of thermal dileptons and photons, and energy losses of
jets.

At high energies the inclusive differential cross sections of hard
processes in nuclear collisions can be computed (in leading twist approximation)
from a factorized form consisting of nuclear
parton distributions and partonic cross sections. At this level all
nuclear effects are contained in the nuclear parton distributions,
which obey the Dokshitzer-Gribov-Lipatov-Altarelli-Parisi (DGLAP)
evolution equations of perturbative QCD (pQCD) \cite{DGLAP}.  At
higher orders in pQCD, absorption of ($1/\varepsilon$)
singularities into definitions of parton distributions is 
scheme-dependent (usually $\overline{MS}$), which retains the
same universality properties 
as in the case of hard processes for free nucleon scatterings.
Consequently, the same nuclear parton distributions can be used to
compute different hard scattering cross sections in nuclear
collisions. This is the obvious motivation to perform a consistent
DGLAP analysis of nuclear parton distributions, as done in
\cite{KJE,EKR98,EKS98}.

Symbolically, the inclusive hard scattering cross sections for
producing a particle $c$ in a collision of nuclei $A$ and $B$ can be
written as
\begin{eqnarray}
& \displaystyle 
   d\sigma(Q^2,\sqrt s)_{AB\rightarrow c+X} =
   \sum_{i,j=q,\bar q,g} 
   \bigg[ Z_Af_i^{p/A}(x_1,Q^2)+(A-Z_A)f_i^{n/A}(x_1,Q^2)\bigg] 
   \otimes \nonumber  \\
 & \otimes\bigg[ Z_Bf_j^{p/B}(x_2,Q^2)+(B-Z_B)f_j^{n/B}(x_2,Q^2)\bigg]
   \otimes d\hat \sigma(Q^2,x_1,x_2)_{ij\rightarrow c+x}
\label{hardAA}
\end{eqnarray}
where $\hat \sigma(Q^2,x_1,x_2)_{ij\rightarrow c+x}$ is the
perturbatively calculable cross section at a large momentum (or mass)
scale $Q\gg \Lambda_{QCD}\sim200$ MeV, $x_{1,2}\sim Q/\sqrt s$ are
the fractional momenta, $f_i^{p(n)/A}(x_1,Q^2)$ is the
distribution of parton species $i$ in a proton (neutron) of the
nucleus $A$, and correspondingly $f_j^{p(n)/B}$ is that for the
nucleus $B$.  The number of protons in $A(B)$ is denoted by $Z_A
(Z_B)$. For isoscalar nuclei, the parton distributions of bound
neutrons are obtained through isospin symmetry (as in the case of
unbound nucleons), $f_{u(\bar u)}^{n/A}=f_{d(\bar d)}^{p/A}$ and
$f_{d(\bar d)}^{n/A}=f_{u(\bar u)}^{p/A}$.  This is expected to be a
good approximation for non-isoscalar nuclei as well.  Therefore, one
may formulate the studies of nuclear parton distributions simply
in terms of those in bound protons, which are denoted here as
$f_i^{p/A}\equiv f_{i/A}$. It is convenient to define the
nuclear effects in parton distributions in terms of the ratio of the
distribution of the parton species $i$ in a bound proton to that in a
free proton,
\begin{equation}
R_i^A(x,Q^2)\equiv {f_i^{p/A}(x,Q^2)\over f_{i/p}(x,Q^2)}.
\label{eqratios}
\end{equation}

Information on the nuclear parton distributions is primarily obtained
by deep inelastic lepton-nucleus scattering (DIS)
\cite{EMC}-\cite{ARNEODO} and by Drell-Yan (DY) dilepton
production in proton-nucleus collisions \cite{E772,E866}.
The nuclear structure functions $F_2^A(x,Q^2)$, measured in DIS
\cite{EMC}-\cite{ARNEODO} differ from those in
free nucleons.  The ratios $R_{F_2}^A\equiv
\frac{1}{A}F_2^A/\frac{1}{2}F_2^{\rm D}$, where deuterium D
approximates an average free nucleon, show clear and systematic
deviations from unity in various regions of Bjorken-$x$: shadowing
($R_{F_2}^A\le 1$) at $x \lsim 0.1$, anti-shadowing ($R_{F_2}^A\ge 1$)
at $0.1 \lsim x \lsim 0.3$, EMC effect ($R_{F_2}^A\le 1$) at $0.3
\lsim x\lsim 0.7$, and Fermi motion ($R_{F_2}^A\ge 1$) as
$x\rightarrow1$ and beyond.  The New Muon
Collaboration (NMC) high-precision
measurements of the $F_2$ structure function ratios for tin vs.
carbon, $F_2^{\rm Sn}/F_2^{\rm C}$ \cite{NMC96-2}
have also revealed a dependence on
the virtuality scale $Q^2$ at small values of $x$.  These measured
modifications of nuclear structure functions directly imply
modifications of parton distributions in bound nucleons. 

Just as in the
QCD-improved parton model (in lowest order) the structure functions
can be written in terms of parton distributions,
\begin{equation}
F_2^A(x,Q^2)=\sum_q e_q^2\{ 
    Z[xf_q^{p/A}(x,Q^2)+xf_{\bar q}^{p/A}(x,Q^2)]
+ (A-Z)[xf_q^{n/A}(x,Q^2)+xf_{\bar q}^{n/A}(x,Q^2)]\},
\end{equation}
where $q$ is the quark flavour and $e_q$ is the corresponding charge.
In the DGLAP analysis of nuclear parton distributions
\cite{KJE,EKR98,EKS98} it is assumed that the distributions
$f_i^{p/A}$ are factorizable at a sufficiently large initial scale,
$Q_0\gg\Lambda_{QCD}$. Once the input distributions are given at
$Q_0^2$ and at $x \ge x_{min}$, their evolution is predicted by the
DGLAP equations at $Q\ge Q_0^2$ and $x \ge x_ {min}$.  As a result,
the ratios $R_i^A(x,Q^2)$ depend both on $x$ and $Q^2$.  In analogy
with the global analyses of the free parton distributions, the key
problem is then to determine the (nonperturbative) initial
distributions $f_i^{p/A}(x,Q_0^2)$. To constrain these, further
information is needed. This is provided by the DIS measurements
mentioned above (NMC \cite{NMC95-1}-\cite{NMC96-2}, SLAC \cite{SLAC},
E665 \cite{E665-1,E665-2}) and by the Drell-Yan data from the E772
and E866 collaborations in p+$A$ collisions \cite{E772,E866}. In
addition, conservation of momentum and baryon number serve as further
constraints.  We emphasize that the measured $Q^2$ dependence of
the ratio $F_2^{\rm Sn}/F_2^{\rm C}$ \cite{NMC96-2} is also reproduced very
well by the DGLAP evolution \cite{EKR98}.

In the DGLAP analysis \cite{EKR98} of the nuclear parton distributions
the nuclear effects were expressed in terms of free parton distributions
which were assumed to be known, i.e. obtained from a set of
distributions such as CTEQ, GRV, MRS etc. The absolute distributions
from different sets of free parton densities may differ by a fairly
large factor\footnote{ideally of course, there would be only one best
set $ \{f_i(x,Q^2)\} $} and, consequently, these differencies will be
reflected in the absolute nuclear parton distributions as well.  The
ratios $R_i^A(x,Q^2)$, however, vary only within a few percent from
set to set, as shown in \cite{EKS98}. Therefore, for computing hard
processes in nuclear collisions with nuclear effects in the parton
distributions, it is a good approximation to use universal ratios
$R_i^A(x,Q^2)$ which are independent of specific free parton
densities.  A parametrization of $R_i^A(x,Q^2)$, ``EKS98'', was
prepared in \cite{EKS98} for general use, and it is available in
\cite{EKS98loc} and now also in the the CERN PDFLIB library of parton
densities \cite{PDFLIB}.

Some uncertainties, however, remain in the determination of the
nonperturbative input distributions at $Q_0^2$. In this paper, we will
focus on constraining uncertainties in the input sea quark
distributions in the region $x\gsim0.2..0.3$, i.e. approaching the region
of the EMC effect, where the ratios $R_{F_2}^A$ measured in DIS are
dominated by valence quarks. Our goal here is to study to what
extent the NA50 Drell-Yan data for p+p and
p+$A$ collisions at $E_{\rm lab}=450$ GeV and Pb+Pb collisions at
$E_{\rm lab}=158 \,A$GeV at the CERN-SPS \cite{SPSexpDY} can be used to
constrain the EMC effect for the input distributions of the nuclear
sea.

\section{General properties of nuclear corrections}

Let us first discuss in some detail how available data and sum
rules constrain the input nuclear parton distributions, or equivalently the
nuclear modifications $R_i^A(x,Q_0^2)$, in different regions of $x$ in
the DGLAP approach \cite{EKR98}.  Since the DGLAP analysis a
perturbative, the scale evolution must be limited to the region
$Q\ge Q_0\sim 1$ GeV.\footnote{In \protect\cite{EKR98} $Q_0=1.5$ GeV
was chosen.}  Some hints are given, however, by DIS
measurements in the non-perturbative region $Q<Q_0$.  As illustrated
in Fig. 1 of \cite{EKR98}, the experimental constraints from DIS and
DY are not given along a fixed value of $Q^2$ -- as would be
preferable for the DGLAP initial conditions -- but in certain
kinematically correlated regions of $x$ and $Q^2$. Furthermore the
data from DIS and DY are typically in distinct kinematical
regions. For these reasons the input distributions must be
constrained by using a recursive procedure similar to the global
analyses of free parton distributions \cite{CTEQ94,MRS98}.

In first approximation the input nuclear effects for valence and sea
quarks can be assumed to be separately flavor-independent: $R_{u_V}^A(x,Q_0^2)\approx
R_{d_V}^A(x,Q_0^2) \approx R_V^A(x,Q_0^2)$, and 
$R_{\bar u}^A(x,Q_0^2)\approx R_{\bar d}^A(x,Q_0^2) \approx
R_{s}^A(x,Q_0^2)=R_S(x,Q_0^2)$ \cite{EKR98}. 
Thus only three independent input ratios must be constrained at $Q_0^2$:
$R_V^A, R_S^A$ and $R_G^A$.

\subsection*{Quarks and antiquarks}

\begin{itemize}
\item At { $x\gsim 0.3$} the structure function $F_2^A$ is dominated
by valence quark distributions.  The DIS data for $R_{F_2}^A$
therefore only constrains the magnitude of the EMC effect and the
Fermi-motion in $R_V^A$ but not those in $R_S^A$ or in $R_G^A$.

\item At { $0.04\lsim x \lsim 0.3$} the DIS and DY data 
give bounds for $R_S^A$ and $R_V^A$ but in different regions of $Q^2$, 
(see Fig. 1 of \cite{EKR98}).

\item At { $x\lsim0.04$} there are DIS data for the ratio $R_{F_2}^A$
available down to $x\sim 5\cdot10^{-3}$ in the region $Q\gsim1$ GeV
relevant for the DGLAP analysis.  Conservation of baryon number forces
the nuclear valence quarks to be less shadowed than the sea quarks.

\item In the DIS data for $R_{F_2}^A$ at {\bf $x\lsim 5\cdot10^{-3}$}
one enters the nonperturbative region $Q\lsim1$ GeV.  A saturation
behaviour of $R_{F_2}$ in $x \rightarrow 0$ is observed along the
experimentally probed values of $Q^2$ \cite{NMC95-2,E665-1}.  Provided
that the sign of the slope of the $Q^2$ dependence of $R_{F_2}^A$ in
the nonperturbative region remains the same (positive) as what is
measured at $x\sim 0.01$ in the perturbative region
\cite{NMC96-2}, a saturation behaviour, i.e. a weak dependence of
$R_{F_2}^A$ on $x$, can also be expected at $Q_0^2$.  Constraints are
then given by the DIS data in the non-perturbative region, in the sense
that the data give a lower bound for $R_{F_2}^A(x,Q_0^2)$.
Since at small values of $x$ the sea quark distributions
dominate over the valence distributions, $R_S^A$ is also constrained
by the DIS data while the shadowing in $R_V^A$ is restricted by baryon
number conservation.
\end{itemize}

\subsection*{Gluons and sea quarks}

\begin{itemize}
\item The scale dependence of $R_{F_2}^A$ at small values of $x$ is
directly connected with shadowing of gluons: the more deeply gluons
are shadowed, the slower is the evolution of $R_{F_2}^A$.  The ratio
$R_G^A$, can thus be constrained by the measured $Q^2$ dependence of
$F_2^A$ as done in \cite{GP,EKR98}.  Since the $Q^2$ dependence is not
very strong, high-precision data is needed.  In practice only the NMC
data for $Q^2$ evolution of $F_2^{\rm Sn}/F_2^{\rm C}$
\cite{NMC96-2} can be used for constraining the input nuclear gluons
at $0.02\lsim x\lsim0.2$.

\item At $x\lsim0.02$ it can be assumed that 
$R_G^A(x,Q_0^2)\approx R_{F_2}^A(x,Q_0^2)$ for $x\ll1$. This remains
true within about 5 \% even after the DGLAP evolution from $Q_0\sim 1$
GeV to $Q\sim100$ GeV \cite{EKR98}.

\item At $x\gsim0.2$ there are no direct experimental constraints
available but conservation of momentum together with arguments for
stable evolution can be used. In this region one is sensitive to the small
tails of the gluon distributions, and the existence of the EMC effect
for nuclear gluons cannot be deduced based on the momentum sum rule
alone. The evolution of gluon distributions are, however, affected
by the valence quark distributions (but not vice versa), so since an
EMC effect exists for the valence quarks, one will be generated for the
gluons as well.  The evolution equations for gluons and sea quarks are
mutually coupled, so an EMC-like depletion will be generated for the
sea quarks through the DGLAP evolution \cite{KJE}.
Then if the nuclear ratios $R_i^A$ do not move
away from their input values very rapidly, it is a plausible first
approximation to have an EMC effect both for the input gluon and sea
quark distributions. For the sea quark modifications, which is
the major subject of this paper, a simple assumption of
$R_S(x\gsim0.3,Q_0^2)\approx R_V(x\gsim0.3,Q_0^2)$ was made in
\cite{EKR98} for the input modifications of the sea quarks.  We now
move on to study the effects of this approximation on
the Drell-Yan dilepton cross
sections in nuclear collisions at SPS energies.

\end{itemize}

\section{Nuclear effects in DY production}

Nuclear effects in Drell-Yan dilepton production can be divided
into two classes: first, there are ``genuine'' nuclear effects arising
from the dynamics of the nuclear collision. These include nuclear
modifications of parton distributions. Second, even without any
nuclear effects in the parton densities or in the collision dynamics,
the DY cross sections in p+$A$ collisions (normalized per $A$) differ between
isoscalar and non-isoscalar nuclei due to the different relative
numbers of protons and neutrons. It is often convenient to use
deuterium D as a reference, since it is approximately a sum of a
free proton and a free neutron. Then any observed deviation of the
DY cross section for p+$A$ in isoscalar nuclei \cite{E772} from that
for p+D can be interpreted directly as a genuine nuclear effect, such
as a nuclear modification in the parton distributions. For
non-isoscalar nuclei, ratios of DY cross sections for p+$A$
to those for p+D or p+p always show additional isospin effects.

For the purposes of comparison of the DY cross sections from p+$A$ with
those from p+D and p+p, we write the isospin symmetric part of the 
parton distributions of the nucleons in a nucleus with $Z$ protons 
separately, 
\begin{equation}
Zf_i^{p/A} + (A-Z)f_i^{n/A} = 
\frac{A}{2}(f_i^{p/A}+f_i^{n/A})
+ (Z-\frac{A}{2})(f_i^{p/A}-f_i^{n/A}).
\end{equation}
The inclusive cross section for the production of the  Drell-Yan
dilepton pairs of invariant mass $M$ and rapidity $y$ in p+$A$ collisions
can then be written in the lowest order as
\begin{eqnarray}
\frac{d\sigma^{pA}_{DY}}{dM^2dy}&=& \frac{A}{2}\frac{8\pi\alpha^2}{9M^2}\frac{M^2}{s}
\bigg\{\sum_{q=u,d,s,...} e_q^2 \bigg[q_1(\bar q_2^{p/A}+\bar q_2^{n/A})
+ \bar q_1(q_2^{p/A}+ q_2^{n/A})\bigg]
+\cr
&&\,\,\,\,\,\,\,\,\,\,\,\,(\frac{2Z}{A}-1)\sum_{q=u,d} e_q^2\bigg[q_1(\bar q_2^{p/A}-\bar q_2^{n/A})
+ \bar q_1(q_2^{p/A}- q_2^{n/A})\bigg]\bigg\},
\label{dsdM2dy}
\end{eqnarray}
where $q_i\equiv f_q(x_i,Q^2)$ with the momentum fractions
$x_{1,2} = \frac{M}{\sqrt s}{\rm e}^{\pm y}$ and a scale choice $Q^2=M^2$.
The ratio of the inclusive Drell-Yan cross section in a p+$A$ collision
vs. that in a p+p or  p+D collision now becomes

\begin{eqnarray}
&\!\!\!\!\!R&\!\!\!\!\!_{DY}^{A/B}(x_2,Q^2) \equiv  \frac
{\frac{1}{A} {d\sigma^{pA}_{DY}}/{dM^2dy}}
{\frac{1}{B} {d\sigma^{pB}_{DY}}/{dM^2dy}}
\nonumber\\
&&\!\!\!\!\!\!\!
 = \{4[u_1(\bar u_2^A+\bar d_2^A)+ \bar u_1(u_2^A+d_2^A)] +
  [d_1(\bar d_2^A+\bar u_2^A) + \bar d_1(d_2^A+u_2^A)] +
4s_1s_2^A +...\}/N_{DY}^{B}  \nonumber\\
&&\!\!\!\!\!\!\!+ (\frac{2Z}{A}-1)
\{4[u_1(\bar u_2^A-\bar d_2^A) + \bar u_1(u_2^A-d_2^A)]+
[d_1(\bar d_2^A-\bar u_2^A) +  \bar d_1(d_2^A-u_2^A)]\}/N_{DY}^{B}
\label{DRELLYAN}
\end{eqnarray}
where the denominator $N_{DY}^{B}$ only contains free parton densities.
For deuterium ($B=D=2$) $N_{DY}^{B}$ is
\begin{equation}
N_{DY}^D = 4[u_1(\bar u_2+\bar d_2)+\bar u_1(u_2+d_2)] +
  [d_1(\bar d_2+\bar u_2) +\bar d_1(d_2+u_2)] + 4s_1s_2 + ...
\end{equation}
and, correspondingly, for the proton ($B=p=1$) it is
\begin{equation}
N_{DY}^p = N_{DY}^D + \sum_{q=u,d} 
4[u_1(\bar u_2-\bar d_2) + \bar u_1(u_2-d_2)]+
[d_1(\bar d_2-\bar u_2) +  \bar d_1(d_2-u_2)].
\end{equation}

It is evident that for isoscalar nuclei, $A = 2Z$, the ratio
$R_{DY}^{A/D}$ is unity in the absence of nuclear modifications
in the parton densities. 
At large
rapidities (large $x_F$) $x_2\ll x_1$ and the ratio $R_{DY}^{A/D}$ is
sensitive mainly to the nuclear effects in the sea quark
distributions. In the following, however, we are interested in the
central rapidities $y\sim0$, so $x_1\sim x_2$ and the ratio
$R_{DY}^{A/D}$ thus reflects the nuclear effects both in the sea quark
and in the valence quark distributions.

\begin{figure}[tb]
\vspace{0cm}
\centerline{\epsfxsize=10cm\epsfbox{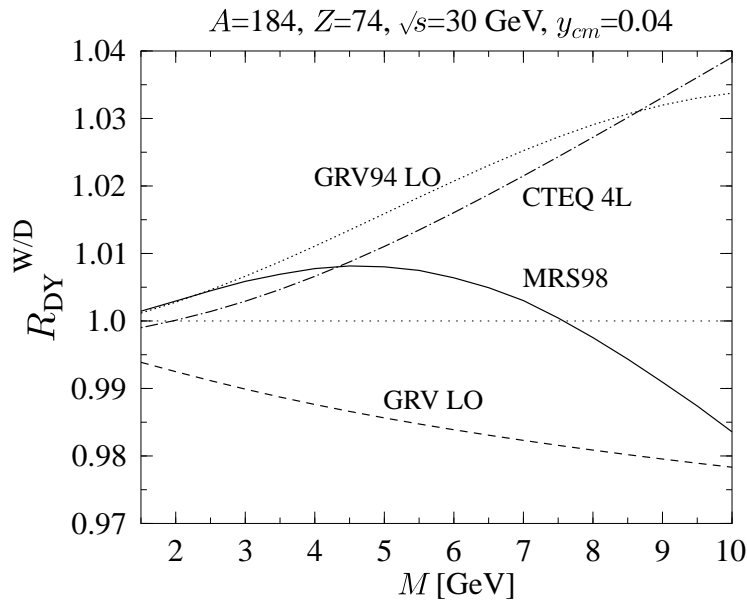}}
\caption[a] { \small Isospin effects in the ratio of DY cross sections
$R_{DY}^{\rm W/D}$ of Eq. (\ref{DRELLYAN}) plotted as a function of mass
$M$ for $^{184}_{~74}{\rm W}$ at $\sqrt s=30$ GeV and $y_{\rm cm}=0.04$,
computed with parton distributions from MRST98 (central gluon) \cite{MRS98} 
(solid), CTEQ4L \cite{CTEQ94} (dotted-dashed), GRV94LO \cite{GRV94} (dotted)
and GRVLO \cite{GRV92} (dashed). Nuclear effects in the parton
distributions are not included.}
\label{isospin}
\end{figure}

For non-isoscalar nuclei on the other hand, even without nuclear
modifications in the parton distribution functions, the isospin
corrections $\sim (\frac{2Z}{A}-1)$ must be considered.  In
fact, they depend quite strongly on the specific set of parton
distributions used in the calculation.  In the oldest sets, $\bar
u=\bar d$ was assumed.  Since $A\ge2Z$, this would lead to
$R_{DY}^{A/D}(x,Q^2)\le1$ in the absence of nuclear effects in the
parton distributions. However, the ratio $\bar u/\bar d$ is
experimentally different from unity: NA51 Collaboration measured this
ratio for the first time, reporting a value of $\bar u/\bar d=0.51\pm
0.04\pm 0.05$ at $x=0.18$ \cite{NA51}. This fact was taken into
account in the subsequent sets of parton distributions, such as MRS94
\cite{MRS94}, GRV94 \cite{GRV94} and CTEQ94 \cite{CTEQ94}. Later on,
E866 Collaboration at Fermilab measured this ratio with higher
accuracy \cite{uoverdFerm}. The modern parton distribution sets now
include the $\bar u/\bar d$ asymmetry. As a result, the behaviour of
the isospin corrections for the Drell-Yan process calculated by using
the recent parton distribution sets differ from those calculated by
using the old ones. To demonstrate this, we plot in Fig. \ref{isospin}
the ratio $R_{DY}^{{\rm W}/D}$ computed with the parton densities of
the free proton. For tungsten, $A=184$ and $Z=74$. The ratio is shown
as a function of mass $M$ at $\sqrt s=30$ GeV and $y_{\rm cm}=0.04$
for four different sets of parton distributions.

%
%

Next, we study how the nuclear effects in the parton distributions and
the isospin effects are reflected in the ratios
$\frac{1}{A}\frac{d\sigma_{DY}^{pA}}{dM}/
\frac{1}{2}\frac{d\sigma_{DY}^{pD}}{dM}$ and
$\frac{1}{A}\frac{d\sigma_{DY}^{pA}}{dM}/
\frac{d\sigma_{DY}^{pp}}{dM}$.
These ratios can be formed from NA50 experimental results. 
They have measured inclusive dilepton
production in p+p, p+D, p+ $_4^9{\rm Be}$ and p+$_{~74}^{184}{\rm
W}$ collisions at $E_{\rm lab}=450$ GeV ($\sqrt s=30$ GeV) at the CERN
SPS in the rapidity range $3<y_{\rm lab}<4$ ($-0.46<y_{\rm cm}<0.54$)
and mass $M$ around the $J/\Psi$ peak. For $M\gsim 4$ GeV, the mass
spectrum is dominated by Drell-Yan dileptons.

For the following calculations, we integrate cross sections (\ref{dsdM2dy}) over the
NA50 rapidity bin, and form the above ratios.
For the free parton distributions we
use the set MRST98 (central gluons) \cite{MRS98} and the nuclear
effects in the parton distributions are taken from the EKS98
parametrization \cite{EKS98}.

%
%
\begin{figure}[tb]
\vspace{0cm}
\centerline{\epsfxsize=10cm\epsfbox{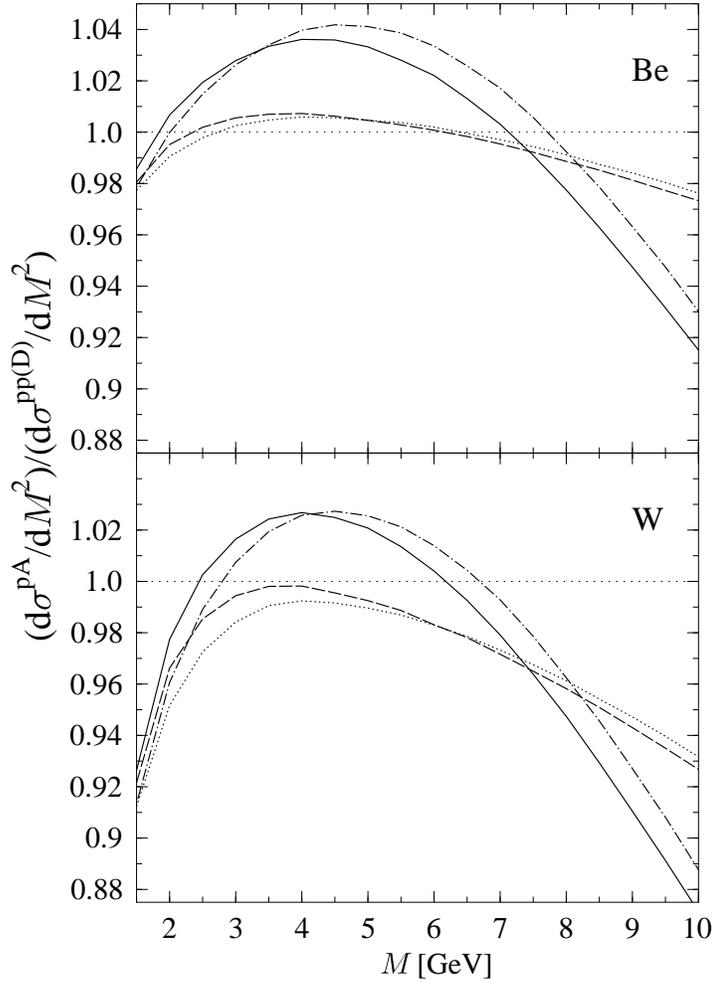}}
\caption[a] { \small The $M$ dependence of the ratios of Drell-Yan
  cross sections $\frac{1}{A}\frac{d\sigma^{pA}}{dM}$ for p+$A$-collisions at
  $\sqrt{s}=30$ GeV and $3 < y_{\rm lab} < 4$.  Upper panel: p+Be over
  p+p for LO (solid) and for NLO (dotted-dashed), p+Be over p+D for LO
  (dashed) and for NLO (dotted). Lower panel is the same for p+W.
  Shadowing and isospin corrections are taken into account }
\label{figBeW2}
\end{figure}

In Fig. \ref{figBeW2} these ratios
are plotted in lowest order for
$^9_4{\rm Be}$ (upper panel; solid, dashed) and $^{184}_{~74}{\rm W}$
(lower panel; solid, dashed).  The analysis \cite{EKR98} for the
nuclear effects $R_i^A(x,Q^2)$ is only a leading order one, so
strictly speaking it should be used only together with the leading
order parton densities. The $Q^2$ evolution of the ratios
$R_i^A(x,Q^2)$, however, is relatively slow, so the ratios given by
EKS98 also serve as a first approximation for the nuclear effects in 
the next-to-leading-order (NLO) parton distributions.  Keeping this
source of uncertainty in mind, we have also computed the Drell-Yan
cross sections in NLO\footnote{Fortran code from P.J. Rijken and W.L. van Neerven is used.} \cite{vanNeerven}. The ratios resulting from the
NLO computation are also shown in Fig.  \ref{figBeW2} for $^9_4{\rm Be}$
(upper panel; dotted-dashed, dotted) and $^{184}_{~74}{\rm W}$ (lower
panel; dotted-dashed, dotted). As seen in the figure, the LO ratios
are a good approximation to the NLO ratios.

In NLO, ${\cal O}(\alpha^2\alpha_s)$, the Drell-Yan cross section
consists of quark-antiquark annihilations with an emission of one real
gluon; $q\bar q_A, \bar q q_A\rightarrow g\gamma^*$, and
gluon-initiated Compton scatterings; $qg_A, gq_A\rightarrow q\gamma^*$
and $\bar q g_A, g\bar q_A\rightarrow \bar q\gamma^*$, and one-loop
corrected quark-antiquark annihilations interfered with the LO
annihilation.  The total NLO contribution to $d\sigma/dM^2dy$ can be
written as a sum $d\sigma_{NLO} = \sigma_{S+V}+\sigma_{Hq\bar
q}+\sigma_{gq}$, where $\sigma_{S+V}$ are the virtual corrections
summed together with the soft emission of a gluon in the $q\bar q$
annihilations, $\sigma_{Hq\bar q}$ contains the emissions of hard
gluons in $q\bar q$ annihilations, and $\sigma_{gq}$ accounts for the
Compton processes \cite{vanNeerven}.  In the kinematical range studied
here, $\sigma_{S+V}$ dominates the net NLO contribution. The sign of
$\sigma_{Hq\bar q}$ changes: $\sigma_{Hq\bar q}/\sigma_{S+V}\sim
+0.04...-0.39$ for $M=10...1.5$ GeV and $y_{\rm cm}=0.04$.  The
Compton term $\sigma_{gq}$ remains negative and
$\sigma_{gq}/\sigma_{S+V}\sim -0.12...-0.09$. Thus the net NLO effect
is always a sum of partly canceling contributions and $K_{DY}= \frac
{d\sigma_{DY}^{NLO}}{dMdy}/\frac {d\sigma_{DY}^{LO}}{dMdy}\sim
1.6... 1.7$.  In the ratios $R_{DY}^{A/D}$ and $R_{DY}^{A/p}$,
however, the $K$-factors largely cancel and the ratios remain very
close to those computed in the LO, as seen in Fig. \ref{figBeW2}.

\begin{figure}[tb]
\centerline{\epsfxsize=10cm\epsfbox{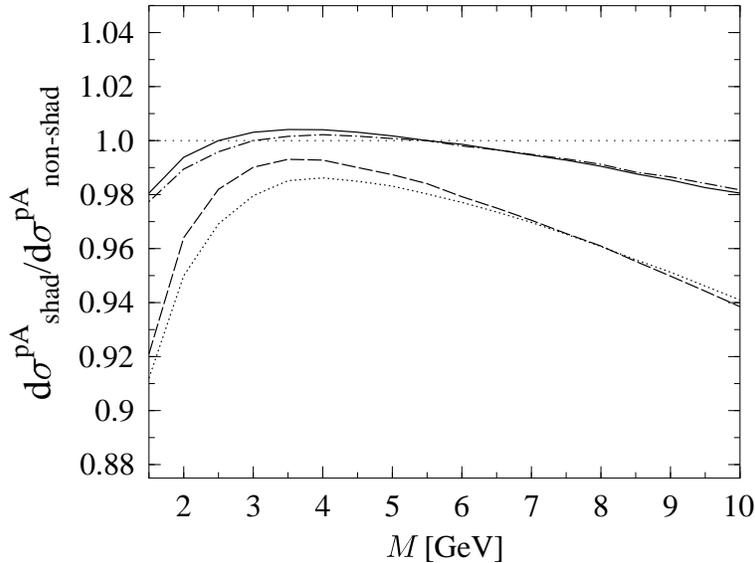}}
\caption[a] { {\small The ratio of shadowed over non-shadowed Drell-Yan
  cross sections $\frac{d\sigma^{pA}}{dM}$ for p+Be LO (solid), p+Be NLO (dotted-dashed), p+W LO
  (dashed) and p+W NLO (dotted) collisions at  $\sqrt{s}=30$ GeV
  and $3 < y_{\rm lab} < 4$. }  }
\label{figpArat}
\end{figure}

The ratios in Fig. \ref{figBeW2} contain both the nuclear effects in
the parton densities and the isospin corrections. To see the effect of
the nuclear parton distributions alone, we have plotted in
Fig. \ref{figpArat} the ratio of the cross sections
$\frac{1}{A}\frac{d\sigma_{DY}^{pA}}{dM}$ computed with and without
nuclear effects in the parton distributions. Both the LO and NLO
ratios are shown. The isospin corrections are taken into account in
all cross sections; thus without nuclear modifications of the parton
densities all the ratios in the figure would reduce to unity.  By
comparison with Fig. \ref{figBeW2}, and from Fig. \ref{figpArat}, we
conclude that the isospin corrections to the ratios $R_{DY}^{A/D}$ remain
small in magnitude when the MRS98 distributions are used.

%
%

\begin{figure}[tb]
\vspace{0cm}
\centerline{\epsfxsize=10cm\epsfbox{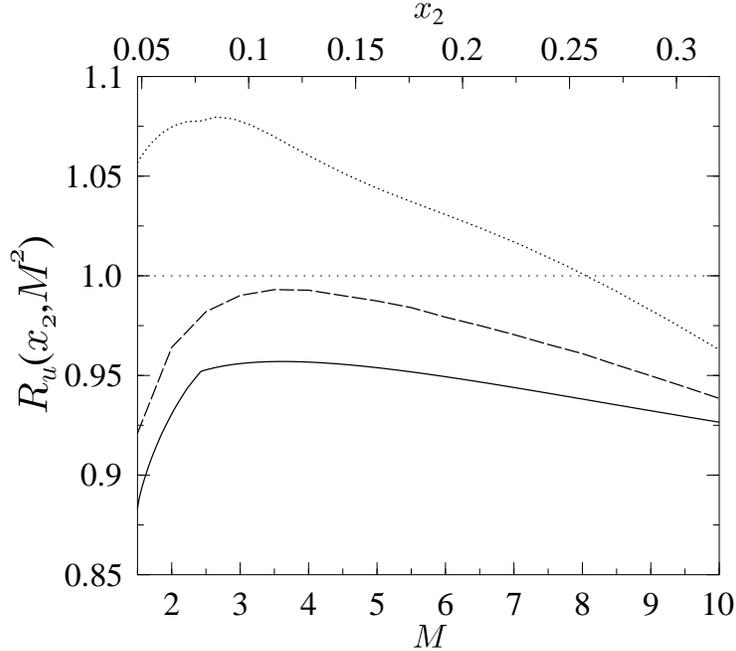}}
\caption[a] { \small The ratios $R^A_{\bar u}(x_2,M^2)$ (solid) and
$R^A_{u_V}(x_2,M^2)$ (dotted) for $^{184}_{~74}$W as a function of
invariant mass $M$ (lower $x$-axis) and $x_2=\frac{M}{\sqrt s}{\rm
e}^{-y_{\rm cm}}$ (upper $x$-axis) for $\sqrt s=30$ GeV and $y_{\rm
cm}=0.04$. The ratio of shadowed vs. non-shadowed LO DY cross sections
(dashed) is the same as the dashed curve in Fig. \ref{figpArat}.
}
\label{figextr}
\end{figure}

Figs. \ref{figBeW2} and \ref{figpArat} indicate that the net nuclear
effects in DY caused by the nuclear modifications of parton densities
are not very dramatic in p+$A$ collisions at the SPS energy 
$E_{\rm lab}=450$ GeV$/c$ in the kinematic region 1 GeV $\lsim M\lsim
10$ GeV, $3<y_{\rm lab}<4$. There are two reasons for this: first, at
the corresponding values of $x_2$ the nuclear effects for the sea
quarks are small. Second, since one is predominantly in the
antishadowing region for valence quarks, there is a cancellation
in the net nuclear effects in the ratio shown. To demonstrate this,
and to show to what extent the ratio p$A$/pD reflects the nuclear
modifications of the sea and valence quarks, we show in
Fig. \ref{figextr} the nuclear effects $R_{u_V}^A(x_2,M^2)$ and
$R_{\bar u}^A(x_2,M^2)$ as a function of mass $M$, with $x_2$ computed
in the middle of the accepted rapidity bin (LO only),
$x_2=\frac{M}{\sqrt s}{\rm e}^{-y_{\rm cm}}$ where $y_{\rm cm}=0.04$. 
The net effect in the DY ratio pW/pD is shown again by the dashed curve.
The values of $x_2$ can be read off from the top of the figure.  Note here
that the scale evolution of the sea quark modifications is taken into
account.

%
%

\begin{figure}[tb]
\centerline{\epsfxsize=10cm\epsfbox{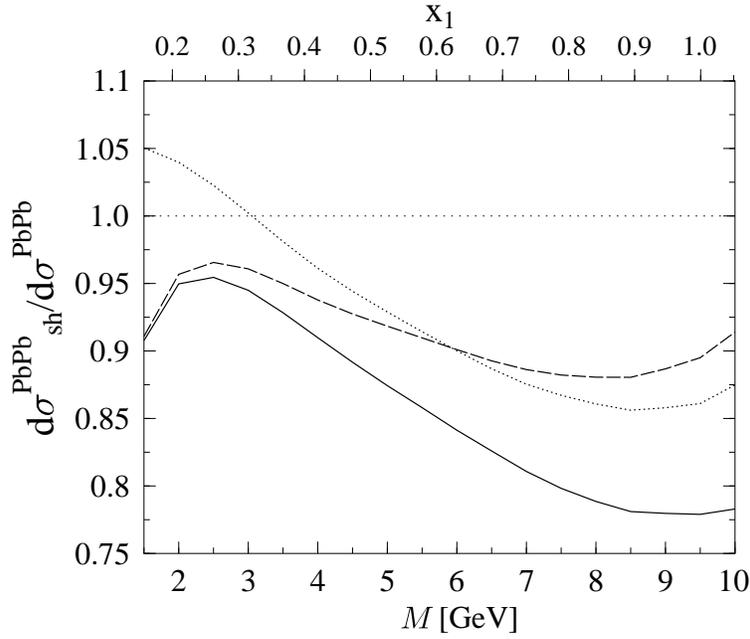}}
\caption[a] { \small The ratio (\protect\ref{eqratPb}) of the
distributions $d\sigma/dM^2$ at $3<y_{\rm lab}<4$ computed with and
without the nuclear effects in parton distributions for Pb+Pb
collisions at $E_{\rm lab}=158\ A$GeV.  Solid curve: $R_{\bar q}^A$
taken from the EKS98 parametrization.  Dashed curve: $R_{\bar q}^A$
interpolated over the region $0.1<x<0.85$ by assuming no EMC
effect. Dotted curve: $R_{\bar q}^A=1$.  Isospin effects are included.
The values of $x_2=\frac{M}{\sqrt s}{\rm e}^{-y_{\rm cm}}$ and
$x_1=\frac{M}{\sqrt s}{\rm e}^{y_{\rm cm}}$, computed in the middle of
the $y$ bin can be read off from the upper $x$ axis.
}
\label{figPbPb}
\end{figure}

To constrain the nuclear effects of the
sea quarks at larger values of $x$, we must consider lower energies.
The NA50 Collaboration at CERN has measured dilepton production in
Pb+Pb collisions at $E_{\rm lab}=158\ A$GeV ($\sqrt s=17.2$ GeV).
This offers us a better chance for constraining the EMC effect in the
input sea quark distributions.
To illustrate the sensitivity of the DY dilepton cross sections
$d\sigma^{\rm Pb+Pb}/dM^2$ (LO, integrated over $3<y_{\rm lab}<4$) to
the assumption of the EMC effect in the input modifications of the
nuclear sea quarks, the ratios 
\begin{equation}
  \frac{ {d\sigma^{\rm Pb+Pb}_{sh}}/{dM^2} }
       { {d\sigma^{\rm Pb+Pb}}/{dM^2} } \label{eqratPb}
\end{equation}
are plotted in Fig. \ref{figPbPb}.  The cross section
$\frac{d\sigma^{\rm Pb+Pb}}{dM^2}$ is the DY dilepton cross
section which includes the isospin effects as in Eq. \ref{hardAA} but
no nuclear effects of the parton distributions.  The cross section
$\frac{d\sigma^{\rm Pb+Pb}_{sh}}{dM^2}$ similarly includes the isospin
effects but is computed by using three different scenarios for the
nuclear effects of the sea quarks. First we take into account all the
nuclear effects as given by the EKS98. The resulting ratio is shown by
the solid curve in Fig. \ref{figPbPb}.  Second, we keep the nuclear
effects of the valence quarks in accordance with EKS98 but relax the
assumption of the EMC-effect in the sea component.  We take the ratios
$R_{\bar q}^A (=R_{q_S}^A)$ from EKS98 at $x<0.1$ (where these ratios are
constrained by experimental data) but interpolate $R_{\bar q}^A$
linearly from $x=0.3$ to the region of Fermi motion $x\gsim0.8$
without an EMC effect\footnote{This procedure would cause some
inconsistency with the EKS98-modification of gluons due to the scale
evolution of the parton densities but as we now do the DY computation
in the LO only, gluons are not involved.}. The resulting ratio is shown
by the dashed curve in Fig. \ref{figPbPb}. Finally, we switch off all
nuclear effects in the sea quarks and antiquarks by setting $R_{\bar
q}^A=1$. We emphasize that the last scenario is actually unphysical as
it violates the available constraints, and that it is meant only for
comparison purposes, to see the effects of the nuclear corrections in
valence quarks alone.  The corresponding ratio is shown by the dotted
curve.  Note that the deviation from unity of the
ratios in Fig. \ref{figPbPb} directly
shows the effects of the nuclear modifications in the parton
distributions. 
The net effect of nuclear parton distributions is now clearly
larger than in the p+$A$ case.

In principle it should be possible to form this ratio
from the measurements, by taking the numerator directly from Pb+Pb
data and the denominator from p+p and p+D data.  To our
knowledge, however, no experimental data presently exists for Drell-Yan
production in p+p or p+D collisions at $E_{\rm lab}=158\ A$GeV$/c$. 
Thus, some additional input is needed in order to form the
experimental ratio (\ref{eqratPb}). One possibility is to compare
the measured DY cross section in
Pb+Pb directly with a purely theoretical
calculation. Alternatively, the denominator in Eq. (\ref{eqratPb})
could be formed from other p+p or p+D data (e.g. data from NA51
collaboration at $E_{\rm lab}=450\ $GeV$/c$ \cite{NA51}) corrected to
$E_{\rm lab}=158\ A$GeV$/c$, based on theoretical cross sections without
nuclear effects in the parton distributions.  In both cases modern
parton distributions, in which constraints from the measured 
ratio $\bar u/\bar d$ are included,
must to be used in order to correctly include isospin corrections.
The main uncertainty would be the overall normalization, which is common to
both cases.
As shown by
Fig. \ref{figPbPb}, the experimental ratio is expected to be 
unity within a $\sim$ 5 \% uncertainty at $M\simeq 4$ GeV. 
If a direct measurement could be made at a single $M$-value, it
could be used to fix the overall normalization of the
ratio. Then the slope of the ratio towards larger values of $M$ would
give direct information of the EMC effect in the sea quark
distributions.

Our study shows that the expected nuclear effects are $\gsim 20$ \% at
masses larger than 7 GeV. Relevant constraints would be obtained for
the EMC effect of the sea quarks and antiquarks if the precision is
$\sim 10$ \%.

%
%

\section{Conclusions}

We have studied the sensitivity of the Drell-Yan cross sections in SPS
nuclear collisions to isospin corrections and to the nuclear
modifications of the parton densities in the regions of antishadowing
and EMC effect. Our aim has been to find a way to constrain the
nuclear modifications of the sea quark distributions in order to
improve the determination of the input modifications in the DGLAP 
analysis \cite{EKR98} of nuclear parton distributions. We have shown 
that the Drell-Yan dilepton data collected by the NA50 collaboration 
at CERN SPS in Pb+Pb collisions at $E_{\rm lab}=158\ A$GeV$/c$ would be 
suitable for constraining the EMC effect in the input distributions 
of the sea quarks provided that a sufficient precision, $\sim 5...10$\%
is reached in forming the ratio $\frac{d\sigma^{\rm Pb+Pb}_{sh}}{dM^2} /
\frac{d\sigma^{\rm Pb+Pb}}{dM^2}$ at $M\gsim5$ GeV.

We have also shown that for p+$A$ collisions at $E_{\rm lab}=450$ GeV, in the
kinematic range $3<y_{\rm lab}<4$, the net effects due to nuclear
modifications of the parton densities are small. This is because
typical values of $x$ remain in the region where the nuclear
effects in the sea quarks and valence quarks largely cancel. Even at
the highest masses studied, $M\sim10$, where the typical $x$ for
$E_{\rm lab}=450$ GeV is $\sim 0.3$, the nuclear parton distributions
modify the p+W cross sections only by 6 \%. This sets the minimum
precision required for such an experiment to constrain the nuclear
sea quark distributions in the antishadowing region. We have also
shown that the isospin effects are small, provided that modern parton 
distributions, where $\bar u\ne\bar d$, are used.

Based on the data from p+$A$ collisions, it is often assumed that
Drell-Yan behaves as $A^1$ in nuclear collisions. From our results for
the SPS, Figs. \ref{figBeW2} and \ref{figpArat}, we see that strictly
speaking this is not the case but the deviations remain fairly small, and
within a 5 \% uncertainty in the cross sections the deviations can be
neglected. However, at higher energies, such as $E_{\rm lab}=800$ GeV in the
Fermilab E772 experiments, the shadowing corrections at smaller values
of $x$ become important and have been experimentally
observed \cite{E772,E866}.

Finally, let us comment on the consequences of these nuclear effects
in the analysis of the Drell-Yan cross sections in Pb+Pb collisions
measured by the NA50 collaboration in connection with $J/\Psi$
suppression \cite{SPSexpDY}. We have shown that the slope of the
invariant mass distributions of the DY pairs is affected by nuclear
effects: Fig. \ref{figPbPb} indicates that corrections of the order of
20 \% appear at $M\sim7..8$ GeV (assuming the EKS98 modifications) but
below the $J/\Psi$ peak they are only about 5\%.  The data points at
large masses, however, have a smaller weight in the $\chi^2$ fits
\cite{SPSexpDY} due to relatively large statistical uncertainties.
The fits are dominated by pair masses near 4 GeV where the error bars
are smaller. The nuclear effects in the mass distributions thus remain
less than 5\% for the extrapolation of the DY cross sections from 4
GeV down to 3 GeV. On the other hand, the experimental $K$-factor,
$K=\sigma_{\rm exp}^{\rm DY}/\sigma_{\rm th}^{\rm DY,LO}$
\cite{SPSexpDY} includes the region of large M, and thus could be
underestimated by 10..15 \% if the nuclear effects in parton
distributions are neglected.

\bigskip

\noindent{\bf Acknowledgements.}  

We thank C. Louren\c co, V. Ruuskanen and Yu.M. Shabelski for useful
discussions.  CAS thanks Ministerio de Educaci\'on y Ciencia of Spain
for financial support. This work was supported by the Academy of
Finland, grant no. 42376.

\end{document}